# Beware of the Small-world neuroscientist!

David Papo[1,*], Massimiliano Zanin[2,3], Johann H. Martínez[4,5], and Javier M. Buldú[1,6]

[1] Laboratory of Biological Networks, Center for Biomedical Technology & GISC, UPM, Madrid, Spain
[2] Faculdade de Ciencias e Tecnologia, Departamento de Engenharia Electrotecnica, Universidade Nova de Lisboa, Lisboa, Portugal
[3] Innaxis Foundation & Research Institute, Madrid, Spain
[4] Universidad del Rosario de Colombia
[5] GISC. Grupo Interdisciplinar de Sistemas Complejos.
[6] Complex Systems Group & GISC, Universidad Rey Juan Carlos, Móstoles, Spain

Characterizing the brain's anatomical and dynamical organization and how this enables it to carry out complex tasks is highly non trivial. While there has long been strong evidence that brain anatomy can be thought of as a complex network at micro as well as macro scales, the use of functional imaging techniques has recently shown that brain dynamics also has a network-like structure.

Network Science [1] allows neuroscientists to quantify the general organizing principles of brain structure and dynamics at all scales in terms of highly reproducible, often universal properties shared by *prima facie* very different systems [2]. A network representation also helps addressing classical but complex issues such as structure-function relationships in a straightforward and elegant fashion, and determining how efficiently a system transfers information or how vulnerable it is to damage [3,4].

One of the most studied global network properties is the *small-world* (SW) structure [5]. In a SW network, nodes tend to form triangles, making the network locally robust. At the same time, the distance between any pair of nodes is much smaller than the network size and increases slowly (logarithmically) with the number of nodes in the network. This combination of properties has been suggested to represent a solution to the trade-off between module independence and specialization, and has been associated with optimal communication efficiency, high-speed and reliability of information transmission [2,3].

In neuroscience, the SW structure has been reported for healthy brain anatomical and functional networks, and deviations from this global organization in various pathologies [6,7]. While there has been some heterogeneity in the adopted definition of SW network, these findings gave the neuroscience community hope that the SW could constitute a functionally meaningful universal feature of global brain organization.

In spite of this preliminary evidence, whether or not the brain is indeed a SW network is still very much an open question [8]. The question that we address here is of a pragmatical rather than an ontological nature: *independently of whether the brain is a SW network or not, to what extent can neuroscientists using standard system-level neuroimaging techniques interpret the SW construct in the context of functional brain networks?*

In a typical experimental setting, neuroscientists record brain images, define nodes and links, construct a network, extract its topological properties, to finally assess their statistical significance and their possible functional meaning. We discuss evidence (some of which is already familiar to the neuroscience community) showing that behind each of these stages lurk fundamental technical, methodological or theoretical stumbling blocks that render the experimental quantification of the SW structure and its interpretation in terms of information processing problematic, questioning its usefulness as a descriptor of global brain organization. The emphasis is on functional brain activity reconstructed using standard system-level brain recording techniques, where the SW construct appears to be the most problematic.

**Small-world property, small-word networks, and small-worldness**

While the SW construct has enjoyed vast popularity in the neuroscience community, a careful look at the literature shows that the various studies resorted to three different though related definitions of SW, which turn out to be nested into each other.

The SW *property* designates networks in which the shortest path $L$ (i.e., the average number of steps needed to go from a node to any other node in the network) is much smaller than the network size $N$ ($L \ll N$) [9]. In a SW *network*, few connecting links drastically shorten the distance between closely knit groups of nodes, so that $L$ is low and grows very slowly with $N$ ($L \sim ln(N)$), while the clustering coefficient $C$ (i.e. the percentage of node's neighbours that are, in turn, linked between them) remains high [5]. Finally, the SW-*ness parameter* $\sigma$ is a continuous, quantitative, measure defined as

$$\sigma = \frac{C/C_{ran}}{L/L_{ran}}$$

i.e. the ratio between $C$ and $L$ normalized by the $L_{ran}$ and $C_{ran}$ of a set of equivalent random networks [10].

In the remainder, attention is mostly focused on $\sigma$, which encompasses the two preceding constructs.

**Pitfalls along the way: from brain recording to data interpretation**

1. *Brain recording devices and standard analyses used to construct networks from neural data can distort the extent to which a network may appear SW.*

The basic ingredients of a SW network are its clustering coefficient $C$ and the average shortest-path $L$. Estimates of these properties crucially depend on the way nodes and links are defined. Different definitions modify $C$ and $L$, ultimately affecting the estimated SW character of the network.

Various sources of possible distortion arising at the first step, that of brain recording, have been illustrated in a number of studies [11-15], including problems due to parcellation and edge definition, spatial embedding, and edge density. For example, in classical electrophysiological methods, nodes are identified with sensors. The lattice-like sensor organization can lead to overestimating the extent to which a network is SW, as different sensors may measure the activity of the same region, ultimately increasing $C$ [14].

Even supposing that brain activity is recorded with an error-free device, projecting brain data onto a network structure comes with its own problems. When dealing with magnetic resonance imaging data, defining nodes is highly non-trivial and may be carried out in different ways, each introducing its own bias, e.g. network reconstruction based on voxel-voxel correlations over-represents connectivity between neighbouring voxels, increasing $C$, whereas parcellations based on different atlases lead to differences in the SW-ness parameter [16].

Estimate distortions also arise from the possible ways of defining links. While there is no well-established criterion to choose a connectivity metric out of the many existing ones, different metrics lead to different connectivity patterns, which may be associated with different basic topological properties, affecting SW evaluation. Moreover, limitations in the reliability of link estimation (e.g. due to noise or common sources) may decrease $L$ and increase $C$, by simply adding a few false positive connections, leading to the observation of SW even in regular or random networks [14]. Furthermore, in its standard formulation, the SW requires networks to be connected, as $d$ diverges in the presence of disconnected nodes. This issue can be dealt with either by adding links (but, this may introduce spurious ones); by taking into account the connected giant component, (but this alters the network size, complicating network comparisons [15]); or by resorting to an equivalent efficiency measure avoiding divergence for disconnected nodes [17].

Furthermore, SW estimates are sensitive to thresholds adopted to prune non-significant links: for high threshold values, brain activity appears hierarchically organized into modules with *large-world* self-similar properties, while adding just a few weak links can make the network non-fractal and *small-world* [18].

2. *Evaluating SW-ness is non-trivial*

Due to the diversity of brain imaging techniques and methodological tools, functional networks may vary in size and link distribution and, as a consequence, in their topological parameters. For this reason, quantifying SW-ness and comparing it across networks requires normalizing $L$ and $C$.

The metric most commonly used to quantify SW-ness, the SW parameter $\sigma$, mainly relies on a normalization using random versions of the original networks [19]. However, how to define an adequate ensemble of random networks is not a straightforward task, as it is unclear what properties of the original network should be conserved. Current methodologies use random rewirings of observed connections, typically conserving the number of nodes and links and the degree distribution, but disregarding the effects of network size on the normalized $C$ and $L$ and the statistical properties of the random ensemble.

The reasons for this standard normalization procedure are to do with the generative model proposed by Watts and Strogatz (WS), which explains the formation of SW as a transition region in a rewiring process from regular to completely random structures [4], making the latter a reasonable reference point. However, the WS mechanism does not reflect the formation of neural connections, suggesting that alternative references, possibly incorporating anatomical or functional constraints, may be more appropriate for normalizing brain networks, and other properties, e.g. link distribution, number of modules in the network, correlations in the number of links, may be conserved in the random versions of observed network structures.

A normalization against random networks may also fail to provide information about the statistical relevance or abnormality of results, an issue that may be dealt with by means of a Z-Score [17]. For instance, two networks with the same normalized clustering coefficient

2.0 may respectively result from a $C$=1.0 and expected $C_{rand}$=0.5, and $C$=0.02 and $C_{rand}$=0.01. While the former network has a clearly abnormal clustering (the highest possible clustering is not to be expected in a random network), the latter may be the result of random fluctuations. Both situations can occur in the same network, as the threshold value above which existing couplings are converted or not into links is varied. Increasing the threshold value induces a shift from a highly clustered network with overabundant links, to one where networks are highly sparse. Notice in addition that, when the overall network size is small, $L$ cannot vary much, so that $\sigma$ values are bound to be strongly correlated with $C$ ones [19]. For functional networks obtained from electro- or magneto-encephalography, $L$ is constrained by the low number of nodes, so that $\sigma$ is mainly controlled by $C$.

*3. The true Aquilles heel of the SW measure lies in interpreting its significance.*

Suppose that the results of unbiased network analyses of brain activity obtained with an ideal recording device point to a SW network. Can this result be taken at face value?

The results of [20] suggest that, for any given degree of disorder $p$, if the system is larger than a crossover size, the network will fall in the SW regime. The percentage $p$ of long-range connections making the network SW scales with the number of nodes $N$ as $N \sim p^{-2/3}$ [21], indicating that only a very small fraction of long-range connections can dramatically decrease $L$. Functional brain imaging studies, which can in principle consider up to $10^5$ nodes, would then typically lead to observing SW-ness.

More importantly, what functional implications should we attribute to a SW brain network? While the SW represents a topological universality class, its functional significance greatly differs in networks of different nature. In communication systems SW networks optimize information processing or transmission efficiency [21], but this is likely not the case for brain networks. The shortest path is usually optimal in a router communication system, whereas in a system such as the brain, other topological (e.g. path redundancy, communicability, branching, loops) and dynamical variables, e.g. burstiness, may better capture information transfer than SW-ness [22-25].

Finally, functional SW networks may result from a diversity of underlying anatomical networks, including randomly connected ones [26]. Thus, the interplay between functional and anatomical networks further enhances the complexity of the SW construct interpretation.

**Concluding remarks: can the SW be salvaged?**

The SW has undeniably been one of the most popular network descriptors in the neuroscience literature. Two main reasons for its lasting popularity are its apparent ease of computation and the intuitions it is thought to provide on how networked systems operate. Over the last few years, some pitfalls of the SW construct and, more generally, of network summary measures, have widely been acknowledged. For instance, analyses using less derivative network measures, such as contrasts of basic edge density have been proposed [27]. However, stress was put far more on technical than on conceptual limitations. The practical advantages of the SW construct often seem to weigh more than fundamental shortcomings. Given the multiple stumbling blocks the SW measure faces, we conclude with two suggestions. First, network normalization should go beyond comparison with "equivalent random networks" and include other properties that account for fundamental properties of brain networks such as modularity, hierarchical structure or spatial embedding. Second, efforts to quantify functional networks' information transfer efficiency or reliability should strive to capture physiologically plausible mechanisms of information transfer and processing. This may involve acknowledging that the universality of network metrics, originally introduced to describe systems profoundly different from the brain, has its limits, and creating a new neuroscience-inspired network science.